\pdfoutput=1
\documentclass{article}
\usepackage[T1]{fontenc}
\usepackage[utf8]{inputenc}
\usepackage{ismir} 
\usepackage{amsmath,amssymb,cite,url,subcaption}
\usepackage{booktabs}
\usepackage{multirow}
\usepackage{graphicx}
\usepackage{color}
\usepackage{placeins}

\title{VIOLET: High-Fidelity Violin Synthesis with Techniques and Dynamics}

\multauthor
  {Baotong Tian$^1$ \hspace{1cm} Cynthia Lu$^1$ \hspace{1cm} Vincent K.M. Cheung$^2$}
  {{\bf  Ting-Kang Wang$^3$\hspace{1cm} Jonathan Churchill$^4$ \hspace{1cm} Zhiyao Duan$^1$}\\
  $^1$ University of Rochester, USA; 
  $^2$ Sony Computer Science Laboratories, Tokyo, Japan\\
  $^3$ National Taiwan University, Taiwan; 
  $^4$ Embertone, USA \\
  {\tt\small \{baotong.tian, zhiyao.duan\}@rochester.edu}
  \vspace{-3mm}
  }
\def\authorname{B. Tian, C. Lu, V.K.M. Cheung, T.-K. Wang, J. Churchill, and Z. Duan}

\usepackage[bookmarks=false,pdfauthor={\authorname},pdfsubject={\pdfsubject},hidelinks]{hyperref}
\usepackage{cleveref}
\usepackage{xurl}

\begin{document}

\maketitle
\begin{abstract}
Neural synthesis for musical instruments has the potential to revolutionize current practices that use concatenative synthesis and a sample library. However, most research focused on piano synthesis and expressive performance generation; little work has been done on continuously articulated instruments like the violin, let alone rendering them with playing techniques and dynamics.
We present \mbox{\textbf{VIOLET}}, a latent-diffusion framework for controllable violin synthesis, which uses a Diffusion Transformer (DiT) with rectified flow to synthesize high-fidelity audio from MIDI notes, playing techniques, and continuous dynamics. To train VIOLET, in addition to using a few existing datasets, we curate a new dataset named \textbf{CSV-TD}, which contains 39~h of 48~kHz synthetic audio and time-aligned annotations of MIDI notes, note-level techniques, and continuous dynamics curves. 
Objective and subjective evaluations show that VIOLET synthesizes violin performances with high technique adherence, accurate pitch and timing alignment, and good dynamics control. It outperforms the current state-of-the-art neural violin synthesis system and approaches a top commercial virtual instrument in terms of technique clarity, naturalness, and dynamics following. 
\end{abstract}

\section{Introduction}\label{sec:introduction}

Audio synthesis for musical instruments aims to generate realistic performance audio from symbolic representations such as MusicXML or MIDI. Recent codec-based and transformer-based systems have substantially improved the expressiveness and perceptual fidelity of generated performances \cite{tang2025midivalle}, but this progress has centered largely on piano. Piano is well suited to event-based modeling because much of its expressive variation is specified at note onset through timing, velocity, and pedaling. The availability of large-scale paired score, MIDI, and audio datasets has further made piano training and evaluation practical at scale \cite{hawthorne2019maestro,foscarin2020asap,zhang2022atepp}. 

Violin performance synthesis presents a substantially different challenge. As a bowed-string instrument, each note is continuously articulated with its pitch, amplitude, spectral content, and temporal envelope shaped throughout the duration of the note instead of just the onset \cite{smith1992waveguides}. Violin performance also involves diverse playing techniques that strongly affect articulation and timbre. High-quality synthesis therefore requires control over both continuously varying dynamics and note-level technique. Professional music production nowadays still relies heavily on sample libraries and virtual instruments (VIs). Although these systems can achieve high sound quality, they incur large storage costs and require labor-intensive programming. Users must specify dense keyswitches and control curves, while transitions between techniques are approximated by stitching and interpolating prerecorded material. Prior work on concatenative synthesis has further shown that rapidly varying expressive parameters can introduce audible discontinuities and require substantial post-processing, especially for continuously controlled instruments \cite{wager2017smooth}.

Related progress has also emerged in the broader field of neural audio synthesis toward explicit control of expressive attributes. In singing voice synthesis, technique-controllable systems such as TechSinger~\cite{guo2025techsinger} show that acoustic generation can be steered using explicit technique labels at phoneme-level resolution. In text-to-speech, recent models have also advanced fine-grained controllability. ControlSpeech~\cite{ji2025controlspeech} enables independent control over timbre, content, and speaking style, while spontaneous-style TTS models~\cite{li2024spontaneoustts} show that fine-grained paralinguistic cues and temporal prosodic variation can be modeled controllably. These developments indicate that combining conditioning over discrete attributes with continuous temporal control is effective for expressive audio generation.

Motivated by this perspective, we introduce \mbox{\textbf{VIOLET}}, a latent-diffusion framework for high-fidelity violin synthesis with explicit control over technique and dynamics. We also curate a new dataset, \textbf{CSV-TD} (Controlled Synthetic Violin with Techniques and Dynamics), containing 39~h of 48~kHz violin audio synthesized from MIDI notes and their annotations of playing techniques and continuous dynamics curves. Our model operates in a latent space using a Diffusion Transformer (DiT) backbone~\cite{peebles2023scalable} trained with a rectified-flow objective~\cite{liu2022rectifiedflow,lipman2023flowmatching}. To achieve separate control for different notes, VIOLET represents technique and dynamics as time-aligned local conditioning signals. Experiments show that the proposed system demonstrates accurate pitch and timing rendering, strong dynamics control, and high technique adherence. It outperforms a state-of-the-art neural violin synthesis method and approaches a top commercial virtual instrument in terms of technique clarity, naturalness and dynamics following.
To the best of our knowledge, this is the first neural violin synthesis system to achieve both high audio quality and explicit control over playing techniques and dynamics\footnote{Code, demo page and dataset are available at \href{https://github.com/User-tian/VIOLET}{\url{https://github.com/User-tian/VIOLET}}}.

\section{Related Work}

\subsection{Neural Music Performance Rendering}
Neural music performance rendering has advanced rapidly in recent years, but most high-performing systems remain centered on piano. Recent approaches span CNN- and Transformer-based score-to-audio models~\cite{performancenet2019,deepperformer2022}, DDSP-based synthesis~\cite{wu2022mididdsp,renault2022differentiable}, state-space models~\cite{dallinger2025pianossm}, and the integration of neural codec language models~\cite{tang2025midivalle}, supported by large paired datasets such as MAESTRO~\cite{hawthorne2019maestro} and ATEPP~\cite{zhang2022atepp}.
However, these methods are best matched to instruments whose expressive variation is largely specified at note onsets. For bowed strings, perceptual realism also depends on continuously evolving bow energy, which makes purely event-centric rendering less adequate.
\subsection{Violin and String Instrument Synthesis}
The dominating approaches for violin and other string-instruments synthesis rely on large sample libraries. In academia, this is often formalized as concatenative synthesis, where recorded material is algorithmically selected, stitched, and smoothed to synthesize a new piece~\cite{schwarz2007corpus,maestre2009expressive}. In  music production, commercial virtual instruments (VIs) utilize a similar approach, employing meticulously recorded samples mapped to dense keyswitches and control curves. While commercial VIs currently represent the industry standard for audio quality, creating a sample library is expensive, time-consuming, hence not scalable to diverse timbre and playing techniques~\cite{wu2022mididdsp,rodet2004data}.

To overcome the inflexibility of sample-based methods, parametric approaches have long been of high interest~\cite{zhang2026review}. Early Abstract Digital Sound Synthesis (ADSS) techniques, such as frequency modulation~\cite{chowning1973synthesis} and wavetable synthesis~\cite{horner1993wavetable}, prioritize computational efficiency and intuitive spectral control, but often lack acoustic realism. In contrast, Physical Modeling Synthesis (PMS) targets structural fidelity through digital waveguides~\cite{smith1992waveguides}, mass-interaction models~\cite{leonard2019migen}, efficient modal simulation~\cite{russo2022efficient}, and refined hysteretic bow--string friction~\cite{matusiak2025refined}. However, the parameters of such physical models remain difficult to tune to capture highly realistic performance nuances.

More recently, neural networks have provided new directions for instrument synthesis. Neural parametric models, such as DDSP-style hierarchical performance modeling~\cite{wu2022mididdsp} and waveform synthesis from string-wise MIDI for guitar~\cite{jonason2024guitar}, offer greater expressive flexibility. Most recently, ViolinDiff~\cite{kim2025violindiff} introduced a two-stage diffusion framework that predicts pitch-bend contours before mel-spectrogram synthesis, demonstrating that explicit F0 modeling can improve expressive violin synthesis. However, existing methods still suffer from lower audio quality compared to sample-based libraries and lack explicit, fine-grained control over playing techniques and dynamics.

\subsection{Generative Audio via Flow Matching}
In parallel, generative audio modeling has shifted from raw-waveform synthesis toward latent generative models that scale more effectively to long, high-resolution signals. AudioLDM~\cite{liu2023audioldm} established latent diffusion as a strong paradigm for text-to-audio generation, while Stable Audio~\cite{evans2024fast} showed that latent DiT-based diffusion can support variable-length, long-form, and high-quality generation. Flow-based approaches have since become increasingly compelling for audio generation. Audiobox~\cite{vyas2023audiobox} demonstrated that flow matching can support controllable unified audio generation, and more recent systems such as FlashAudio~\cite{liu2025flashaudio} and TangoFlux~\cite{hung2025tangoflux} showed that flow-based and rectified-flow formulations can produce high-quality audio with much faster sampling. These results motivate our use of a flow matching framework~\cite{lipman2023flowmatching}. It provides a credible route to high-fidelity audio synthesis, and its continuous-time formulation is well suited to the continuously evolving acoustics of violin performance.
\FloatBarrier
\section{Methodology}\label{sec:methodology}

\begin{figure*}[htbp]
  \centering
  \includegraphics[width=\textwidth]{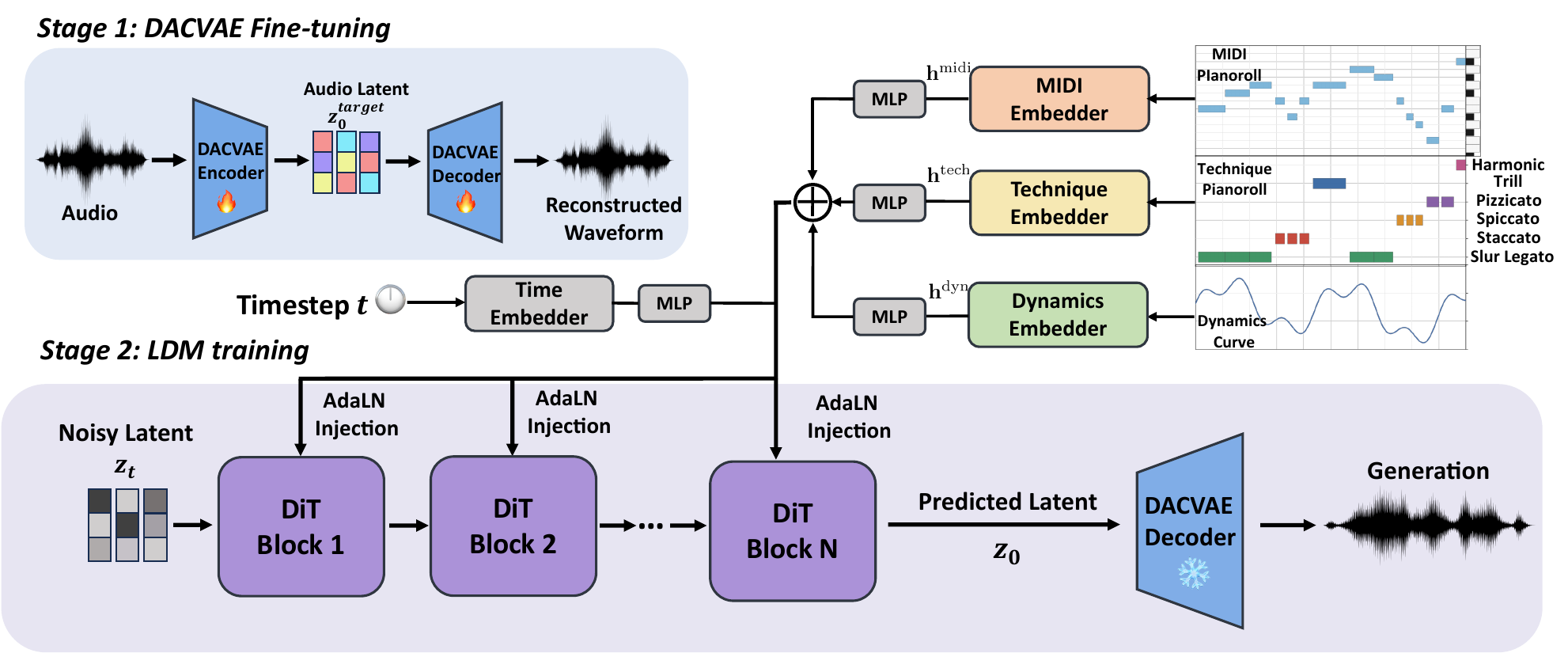}
  \caption{Overall pipeline of the VIOLET framework.}
  \label{fig:violet_pipeline}
  \vspace{-5mm}
\end{figure*}
In this section, we introduce the proposed VIOLET framework for high-fidelity violin synthesis with control over techniques and dynamics. As illustrated in \figref{fig:violet_pipeline}, the system consists of two stages: fine-tuning a DACVAE model to encode violin audio into a compact latent space, and training a Latent Diffusion Model (LDM) to synthesize the latents. Three parallel control signals, namely MIDI notes, playing techniques, and continuous dynamics, are processed by their respective embedders and injected into a Diffusion Transformer (DiT) backbone via Adaptive Layer Normalization (AdaLN) to guide the generation process.
\subsection{Audio Representation}

We adopt DACVAE, the VAE version of the Descript Audio Codec (DAC) \cite{kumar2023high}, as the audio latent representation for violin synthesis. We take the official watermarked checkpoint~\cite{meta2025dacvae} and fine-tune the decoder on violin recordings using the losses from the original training pipeline. After fine-tuning, the codec is frozen: the encoder maps violin audio to latents for diffusion training, and the decoder converts generated latents back to 48~kHz mono audio. The latent frame rate is 25~Hz, corresponding to one representation every 40~ms. Qualitative observations indicate that compared with the original model, the fine-tuned codec improves high-frequency reconstruction for violin, particularly the harmonic fluctuations of vibrato notes. Details regarding the datasets used can be found in \cref{subsec:experimental_details}.
\subsection{Conditioning and Alignment}
\label{sec:cond}
We provide MIDI notes, technique labels and dynamics curves as conditioning inputs to the neural model. Since we need to make sure the audio output follows the timing of the inputs, our conditioning inputs are local time-aligned signals rather than global attributes. This allows note-level technique and continuous dynamics to modulate the latent generation process at each frame. 

For MIDI notes, we represent the condition as a binary pianoroll $R\in\{0,1\}^{P\times T_c}$, where $P$ is the number of semitones in the pitch range and $T_c$ is the number of time frames. We set the pitch range to the playable range of the violin, from G3 to A7. For playing techniques, we construct a similar binary pianoroll $R_{\text{tech}} \in \{0,1\}^{12 \times T_c}$ aligned with the MIDI note pianoroll in time, where each row represents one of the 12 playing techniques assigned at note level. Each technique condition follows the same duration as its corresponding MIDI note. For dynamics, we use the Continuous Controller 1 (CC1) information. In MIDI files, CC1 is typically stored as discrete events with values between 0 and 127; we min-max normalize these values to $[0,1]$ and apply a zero-order hold interpolation, extending each discrete value until the next event to construct a piecewise constant, frame-aligned dynamics curve.

\subsection{Model Architecture}
\label{subsec:model-arch}

After the DACVAE is fine-tuned in Stage 1, we train a DiT-based latent diffusion model in Stage 2.

We map the three conditions described in \cref{sec:cond} into latent sequences with separate embedders before feeding them to the DiT model. The MIDI embedder applies a 1D causal convolution to temporally downsample the note pianoroll to the latent frame rate. This causal design enables the model for real-time use in the future.
The technique embedder directly downsamples the technique pianoroll to the same latent frame rate. The dynamics embedder projects the dynamics curve with a linear layer. 

After passing the representations through the embedders, the resulting three control embeddings, $h^{\mathrm{midi}}$, $h^{\mathrm{tech}}$ and $h^{\mathrm{dyn}}$, all have shape $\mathbb{R}^{T \times D}$, where $T$ is the latent sequence length, and $D$ is the model latent dimension.
This shared representation allows each control to contribute modulation parameters at every latent frame.

We follow the adaptive layer normalization (AdaLN) design from DiT \cite{peebles2023scalable}, where a conditioning MLP predicts per-frame modulation parameters $(\alpha, \beta, \gamma)$ for each module within the transformer block. Specifically, $\gamma$ and $\beta$ determine the scale and shift applied during adaptive layer normalization, while $\alpha$ acts as a gate that scales the contribution of the module output prior to the residual addition. In the standard DiT setting, the MLP maps a global conditioning vector to six $D$-dimensional parameters per latent frame, namely $(\alpha^{\mathrm{msa}}, \beta^{\mathrm{msa}}, \gamma^{\mathrm{msa}})$ for the multi-head self-attention (MSA) module and $(\alpha^{\mathrm{ffn}}, \beta^{\mathrm{ffn}}, \gamma^{\mathrm{ffn}})$ for the feed-forward network (FFN) module.
In our setting, we compute the global modulation parameter from the diffusion-step embedding $t_{\mathrm{emb}} \in \mathbb{R}^{D}$ and augment it with three local modulation parameters computed from the three control embeddings through their MLP heads. Both the global and the local parameters are $D$-dimensional.
The final modulation parameter is the sum of the global and local parameters. Taking the shift parameter $\beta^{\mathrm{msa}}_{\ell,f}$ in the multi-head self-attention module in the $\ell^{\mathrm{th}}$ block as an example:
\begin{equation}
\begin{split}
  \beta^{\mathrm{msa}}_{\ell,f} &= \bar{\beta}^{\mathrm{msa}}_{\ell}(t_{\mathrm{emb}})
    + \tilde{\beta}^{\mathrm{msa}}_{\ell,f}(h^{\mathrm{dyn}}_{f}) \\
  &\quad + \tilde{\beta}^{\mathrm{msa}}_{\ell,f}(h^{\mathrm{midi}}_{f})
    + \tilde{\beta}^{\mathrm{msa}}_{\ell,f}(h^{\mathrm{tech}}_{f}),
\end{split}
\label{eq:local-adaln}
\end{equation}
where $f\in\{1,\ldots,T\}$ indexes the latent frames, and $\bar{\cdot}$ is broadcast from size $D$ to $T\times D$. The parameters $\alpha^{\mathrm{msa}}_{\ell,f}$ and $\gamma^{\mathrm{msa}}_{\ell,f}$ are computed similarly. 
The summed parameters are then applied in the multi-head self-attention module as
\begin{align}
  y^{\mathrm{msa}}_{\ell} &= \mathrm{LN}(x_\ell) \odot (1 + \gamma^{\mathrm{msa}}_{\ell}) + \beta^{\mathrm{msa}}_{\ell}, \label{eq:adaln-msa} \\
  x'_{\ell} &= x_\ell + \alpha^{\mathrm{msa}}_{\ell} \odot \mathrm{MSA}_\ell(y^{\mathrm{msa}}_{\ell}), \nonumber
\end{align}
where \(x_{\ell}\) is the input to the \(\ell\)-th block, $\odot$ denotes element-wise multiplication, and the subscript $f$ is dropped for simpler notation. The parameters for the FFN module  $(\alpha^{\mathrm{ffn}}, \beta^{\mathrm{ffn}}, \gamma^{\mathrm{ffn}})$ are computed and used similarly. All modulation head output layers are zero-initialized following the AdaLN-Zero practice \cite{peebles2023scalable}.

\subsection{Training and Inference}

\noindent\textbf{Rectified flow objective.} 
Let $z_0$ denote clean audio latents, $z_1 \sim \mathcal{N}(0,I)$ denote the Gaussian noise, and $c$ represent the combined conditioning signals (MIDI, technique, and dynamics). We sample a flow time $t \in [0,1]$ and form the linear interpolation
\begin{equation}
  z_t \;=\; (1-t)\,z_0 + t\,z_1,
\end{equation}
whose pathwise velocity is constant $v^\star = z_1 - z_0$.
The model $v_\theta(z_t, t, c)$ is trained to predict this velocity via
\begin{equation}
  \mathcal{L} \;=\; \mathbb{E}_{z_0,\,z_1,\,t}\!\Big[
    \big\lVert v_\theta(z_t, t, c) - (z_1 - z_0) \big\rVert_2^2
  \Big].
  \label{eq:rf-loss}
\end{equation}
To enable compositional classifier-free guidance at inference, we randomly replace each conditioning modality with a learned null embedding during training, which allows the model to estimate velocity fields under different subsets of the conditioning signals.

\noindent\textbf{Inference.}
Sampling starts from Gaussian noise at $t=1$ and integrates the learned velocity field backward to clean latents at $t=0$ using discrete Euler steps:
\begin{equation}
    z_{t_{i+1}} \; = \; z_{t_i} + (t_{i+1} - t_i) \, v_\theta(z_{t_i}, \, t_i, \, c),
\end{equation}
where $t_i$ represents the sequence of discretized time steps that decrease from $1$ to $0$. We use a \emph{compositional} classifier-free guidance~\cite{tsai2025musecontrollite} scheme. At each sampling step, we evaluate the same network under three nested condition sets: MIDI only ($v_{\mathrm{m}}$), MIDI and technique ($v_{\mathrm{m,t}}$), and all controls ($v_{\mathrm{full}}$). The guided velocity is the sum
\begin{equation}
  v_{\mathrm{cfg}} \;=\; v_{\mathrm{m}}
    + w_{\mathrm{tech}}\,(v_{\mathrm{m,t}} - v_{\mathrm{m}})
    + w_{\mathrm{dyn}}\,(v_{\mathrm{full}} - v_{\mathrm{m,t}}),
  \label{eq:cfg}
\end{equation}
where $w_{\mathrm{tech}}$ and $w_{\mathrm{dyn}}$ scale the technique and dynamics guidance directions, while MIDI conditioning remains active in all branches. Here we model dynamics as dependent on technique because the two conditions jointly describe acoustic features of violin playing, and dynamics are expressed differently across techniques.

To render durations beyond the model's fixed context window, we divide the target timeline into windows with 50\% overlap, each aligned to the corresponding slice of the MIDI, technique, and dynamics conditions.
Each window is denoised independently, and the resulting waveforms are overlap-added in the time domain using Hann windows to ensure smooth transitions across segment boundaries.

\section{Datasets}
\label{sec:dataset}
To the best of our knowledge, no public violin dataset provides aligned MIDI notes, note-level techniques, and continuous dynamics controls required by our task. We therefore construct \textbf{CSV-TD} with a commercial virtual instrument, obtaining high-quality audio with the exact symbolic controls used for rendering. This section describes CSV-TD and the additional corpora used in this work.

\subsection{CSV-TD Dataset}
We use MID\_FiLD~\cite{ryu2024mid} as the MIDI source material as it provides human-written dynamics curves.
We extracted monophonic lines suitable for solo rendering, then inserted technique controls as MIDI keyswitches below the violin range. Labels were assigned
using duration-based probabilistic heuristics: shorter notes were more
likely to receive spiccato, staccato, or pizzicato, whereas longer
notes were more likely to receive legato, trill, or
harmonic. Finally, we developed a JUCE-based offline rendering framework for Kontakt~\cite{nativeinstruments2018kontakt8}
and rendered the annotated MIDI at 48~kHz in stereo using
Joshua Bell Violin~\cite{embertone2024joshuabell}, a commercial
solo-violin virtual instrument. The resulting CSV-TD training set contains 6{,}108 MIDI--audio pairs totaling 35 hours.



\subsection{Additional Corpora}
\begin{table}[t]
  \centering
  \footnotesize
  \setlength{\tabcolsep}{3pt}
  \resizebox{\columnwidth}{!}{%
  \begin{tabular}{@{}lccccc@{}}
  \toprule
  \textbf{Name} & \textbf{Type} & \textbf{\#} & \textbf{Sample Rate} & \textbf{Annotation} & \textbf{Duration} \\
  \midrule
  CSV-TD train & Syn. & 6{,}108 & 48\,kHz St. & Tech./Dyn. & 35.4~h \\
  CSV-TD test  & Syn. & 686   & 48\,kHz St. & Tech./Dyn. & 3.7~h \\
  MOSA\_VPT    & Syn. & 1{,}864 & 48\,kHz St. & Tech.      & 75.6~h \\
  MOSA         & Real & 461   & 44.1\,kHz M. & N/A      & 18.9~h \\
  MUSC         & Real & 939   & 48\,kHz St. & N/A      & 30.9~h \\
  \bottomrule
  \end{tabular}%
  }
  \caption{Statistics of the datasets used in this work. Syn. refers to synthetic, St. refers to stereo, M. refers to mono.}
  \label{tab:dataset_stats}
  \vspace{-5mm}
\end{table}
In addition to CSV-TD, we use two real violin datasets and one synthetic augmentation dataset for training.
\tabref{tab:dataset_stats} summarizes the corpora used in this work. 

\noindent\textbf{MOSA~\cite{huang2024mosa}.} From this dataset, we retain a filtered violin-only subset containing 19 hours of professional solo violin recordings by 15 expert players. We use the manually aligned MIDI--audio pairs but not the original score-level expressive annotations, which do not directly match our technique taxonomy or dynamics representation.


\noindent\textbf{MUSC~\cite{tamer2023high}.} This dataset consists of solo violin recordings from Wohlfahrt, Kayser, and Paganini etudes. After excluding unavailable recordings due to removed YouTube links, we retain 939 MIDI--audio pairs with a total of 31 hours. The dataset provides aligned MIDI and audio, but no technique or dynamics annotations.

\noindent\textbf{MOSA\_VPT~\cite{wang2026vioptt}.} This dataset is a synthetic augmentation of MOSA with four technique conditions: sustain, harmonic, spiccato, and pizzicato. We use the 48~kHz version with a total of 76 hours of audio. It is used as additional technique-supervised training data.


\section{Experiments}
\subsection{Experimental Setup}
\label{subsec:experimental_details}

\noindent\textbf{Dataset.} We use all the training corpora (two real datasets and two synthetic datasets) to fine-tune the DACVAE model and to train the main latent diffusion model. For objective evaluation, we use the CSV-TD test set. While the CSV-TD training set contains 12 technique labels and we use all of them for training, here we focus on synthesizing and evaluating on 7 common techniques: \textbf{harmonic, pizzicato, slur legato, spiccato, staccato, major trill, and minor trill}. For subjective evaluation, we curated 14 single-technique excerpts, two for each technique, and 3 multi-technique excerpts, each covering a few techniques.

\noindent\textbf{Implementation.} We use a base DiT model as the LDM backbone consisting of 12 DiT blocks with 12 attention heads, and a hidden dimension of 768. The model generates 10~s audio segments at a time. Following \cite{evans2025stable}, we apply Rotary Position Embeddings (RoPE) to half of each attention head dimension and use a gated MLP in each DiT block. During training, we adopt curriculum learning over datasets. The first stage emphasizes synthetic data, with a sampling ratio of CSV-TD : MOSA\_VPT : MOSA : MUSC = \(60\!:\!20\!:\!10\!:\!10\). We then increase the proportion of real recordings to improve natural transitions, overall fidelity, and expressiveness, using a ratio of \(40\!:\!10\!:\!25\!:\!25\). We train the model for 100{,}000 steps on two A100 GPUs with a batch size of 32 and gradient accumulation over 4 batches, which takes approximately 4 days. At inference time, we use a rectified-flow Euler sampler for 30 sampling steps with \(w_{\mathrm{tech}} = w_{\mathrm{dyn}}=1\). On one A100 GPU, generating 10~s of audio takes 2.3~s (i.e., RTF=0.23).

\noindent\textbf{Baselines.} We include ViolinDiff~\cite{kim2025violindiff} as one baseline, which 
is the current state-of-the-art neural synthesis method for the violin. We also include the Joshua Bell Violin as a strong VI reference. To compare different variants of our model, we additionally evaluate \textbf{VIOLET (w/o Cond)}, which renders all notes as sustain notes with constant dynamics to match ViolinDiff's input condition, and \textbf{VIOLET (Synth)}, which is trained only on synthetic datasets (CSV-TD and MOSA\_VPT).

\subsection{Objective Evaluation}
\noindent\textbf{Metrics.}
We evaluate audio quality, MIDI--audio alignment, and dynamics controllability using FAD-48k, onset--pitch F1, onset deviation, and a Spearman correlation with dynamics. For temporal evaluation, directly using the notated MIDI onset can be misleading, since the actual perceived onset is always slightly later than the sample trigger time marked by the MIDI onset in a virtual-instrument rendering pipeline. 
We therefore construct a timing-compensated ground-truth MIDI for VI and all VIOLET systems, shifting each note later heuristically by a technique-dependent delay. 
Following empirical MIDI orchestration practices for virtual instrument pre-delays~\cite{vi_predelay_db} and acoustic studies on bowed string transients~\cite{guettler2002bowed}, we use 30~ms for short articulations including pizzicato, staccato, and spiccato, and 100~ms for longer articulations including slur legato, harmonic, and trills. For ViolinDiff, we keep the original MIDI timing unchanged, since it was not trained on virtual-instrument-rendered data.

For perceptual distribution matching, we compute 48~kHz Fréchet Audio Distance (FAD)~\cite{kilgour19_interspeech}, 
using the fadtk toolkit with the L-CLAP music model~\cite{gui2024adapting,wu2023large}. We upsample the audio generated by ViolinDiff to 48~kHz before the embedding extraction. 
We maintain a reference set consisting of $\sim$34 hours built from MOSA and MUSC datasets, with approximately 17 hours sampled from each corpus.

For MIDI--audio alignment, we use VioPTT \cite{wang2026vioptt} to transcribe synthesized audio into MIDI note events. We report onset--pitch F1 with 50~ms and 100~ms tolerances using \texttt{mir\_eval}~\cite{raffel2014mireval}. Trill notes are excluded from this evaluation because dense ornamental re-articulations can produce multiple detected onsets, whereas our MIDI annotations represent each trill as a sustained base note with a trill technique label. To complement F1 with a direct timing-error measure, we compute the mean absolute onset deviation over matched predicted--reference pairs, measuring how accurately each system places note attacks.

For dynamics evaluation, we calculate Spearman rank correlation (\(\rho\)) between the synthesized audio Root Mean Square (RMS) in dB and the average dynamics values. We compute RMS and average dynamics for each note. However, for notes longer than 1~s with an internal normalized dynamics range above 0.1, we divide them into 1~s segments and compute the RMS and average dynamics for each segment, since crescendo and diminuendo are likely to be prominent within these notes. 
This is calculated for each file and then averaged over the CSV-TD test set. 

\begin{table}[t] 
  \centering
  \setlength{\tabcolsep}{4pt} 
  \resizebox{\columnwidth}{!}{%
    \begin{tabular}{l c c c c}
    \toprule
    \multirow{2}{*}{\textbf{System}} & \textbf{FAD} & \textbf{Onset--Pitch F1} & \textbf{Onset Dev.} & \textbf{Dyn. $\rho$} \\
          & ($\downarrow$) & 50 / 100 ms ($\uparrow$) & 50 / 100 ms ($\downarrow$) & ($\uparrow$) \\
    \midrule
    ViolinDiff & 0.668 & 0.793 / 0.833 & 17.8 / 20.1 & 0.036 \\
    VIOLET (w/o Cond) & 0.526 & 0.722 / \textbf{0.894} & \textbf{14.1} / 23.8 & 0.025 \\
    VIOLET (Synth) & \textbf{0.510} & \underline{0.797} / 0.849 & \underline{14.9} / \textbf{18.0} & \underline{0.620} \\
    VIOLET (Full) & \underline{0.513} & \textbf{0.821} / \underline{0.879} & \underline{14.9} / \underline{18.6} & \textbf{0.631} \\
    \midrule
    VI & 0.428 & 0.884 / 0.925 & 14.7 / 17.1 & 0.671 \\
    \bottomrule
    \end{tabular}%
    }
  \caption{Objective evaluation results. The best and second-best results among the generative models are highlighted in \textbf{bold} and \underline{underlined}, respectively.}
  \label{tab:objective_results}
  \vspace{-5mm}
\end{table}
\noindent\textbf{Results.} \Cref{tab:objective_results} summarizes the objective evaluation results. We can see that VIOLET improves over ViolinDiff while approaching the VI reference. In terms of audio quality, VIOLET achieves a substantially lower FAD than ViolinDiff, indicating better distributional similarity to real violin recordings. This improvement is obtained without sacrificing MIDI--audio alignment, as VIOLET (Full) achieves the best 50~ms onset--pitch F1 among neural systems, and its onset deviation remains close to that of the VI reference. These results suggest that adding technique and dynamics conditioning does not affect the MIDI-following ability of the system.

The benefit of explicit conditioning is most evident in the note-level Spearman correlation metric. VIOLET (w/o Cond) shows little correspondence between rendered RMS and the input dynamics curve, whereas both conditioned variants achieve a high correlation score. This confirms that the proposed model uses the dynamics control to shape continuous loudness variation, rather than only improving overall timbre or fidelity. Finally, VIOLET (Synth) performs similarly to VIOLET (Full) across the objective metrics, suggesting that the synthetic training data already provide strong supervision for controllable violin rendering. The slight improvement in dynamics and onset--pitch F1 suggests that adding real violin training data could be a promising direction for future improvements in expressiveness. However, this benefit is not yet fully pronounced, probably due to the limited scale and noisier recording conditions of the current real violin corpora. Moreover, because the test set is synthetic, the evaluation may favor models trained primarily on synthetic data. These results should therefore be interpreted as measures of basic rendering correctness, particularly adherence to the input MIDI timing and pitch, rather than strong evidence of improved naturalness or generalization to real performances.

\subsection{Subjective Evaluation}
\begin{table}[t]
\centering
\resizebox{\columnwidth}{!}{%
\begin{tabular}{lcccccc}
\toprule
\textbf{System} & \textbf{Slur Leg.} & \textbf{Harm.} & \textbf{Trill} & \textbf{Pizz.} & \textbf{Spicc.} & \textbf{Stacc.} \\
\midrule
VIOLET (Full) & 100.0 & 100.0 & 100.0 & 92.9 & 80.0 & 95.7 \\
VI          & 100.0 & 100.0 & 100.0 & 86.2 & 83.3 & 85.2 \\
\bottomrule
\end{tabular}%
}
\caption{Single-technique identification accuracy (\%) for VIOLET and the VI baseline. Results for major and minor trills are aggregated into a single \textbf{Trill} category.}
\label{tab:technique_accuracy}
\vspace{-5mm}
\end{table}
\noindent\textbf{Experimental Setup.}
The subjective study evaluated perceptual synthesis quality in single- and multi-technique settings using excerpts selected from beginner-to-intermediate violin etude books and repertoire to cover diverse playing techniques. In the single-technique setting, we selected two $\sim$10~s excerpts for each technique and compared VIOLET (Full) with VI, excluding ViolinDiff as it lacks technique conditioning. Listeners identified the perceived technique from a candidate list and rated technique clarity, naturalness, audio quality, and dynamics matching on a 1--5 Likert scale. For dynamics matching, each rendering is conditioned on one of four normalized curves: linear rise, linear fall, arch-shaped rise--fall, or valley-shaped fall--rise. In the multi-technique setting, we used three $\sim$30~s excerpts containing multiple techniques, with dynamics derived from score markings. Given the technique list for each excerpt, listeners rated renderings from VIOLET (Full), VI, and ViolinDiff for technique clarity, naturalness, and audio quality on the same scale.


A total of 15 listeners participated in the study. All self-reported being either professional musicians or highly familiar with violin playing techniques. Each listener was asked to evaluate all renderings in randomized order.

\noindent\textbf{Results.} \Cref{tab:technique_accuracy} summarizes single-technique identification accuracy. VIOLET achieves 100\% accuracy on long-note techniques and remains competitive with VI on short-note articulations. It outperforms VI on pizzicato and staccato, while performing slightly worse on spiccato. Further investigation of the result (not shown in the table) shows that most errors on spiccato identification were confused as staccato, as both are short-note attacks. 
Overall, the results indicate that the proposed conditioning strategy of VIOLET effectively renders the intended playing techniques.

\begin{figure}[t]
  \centering
  \includegraphics[width=\columnwidth]{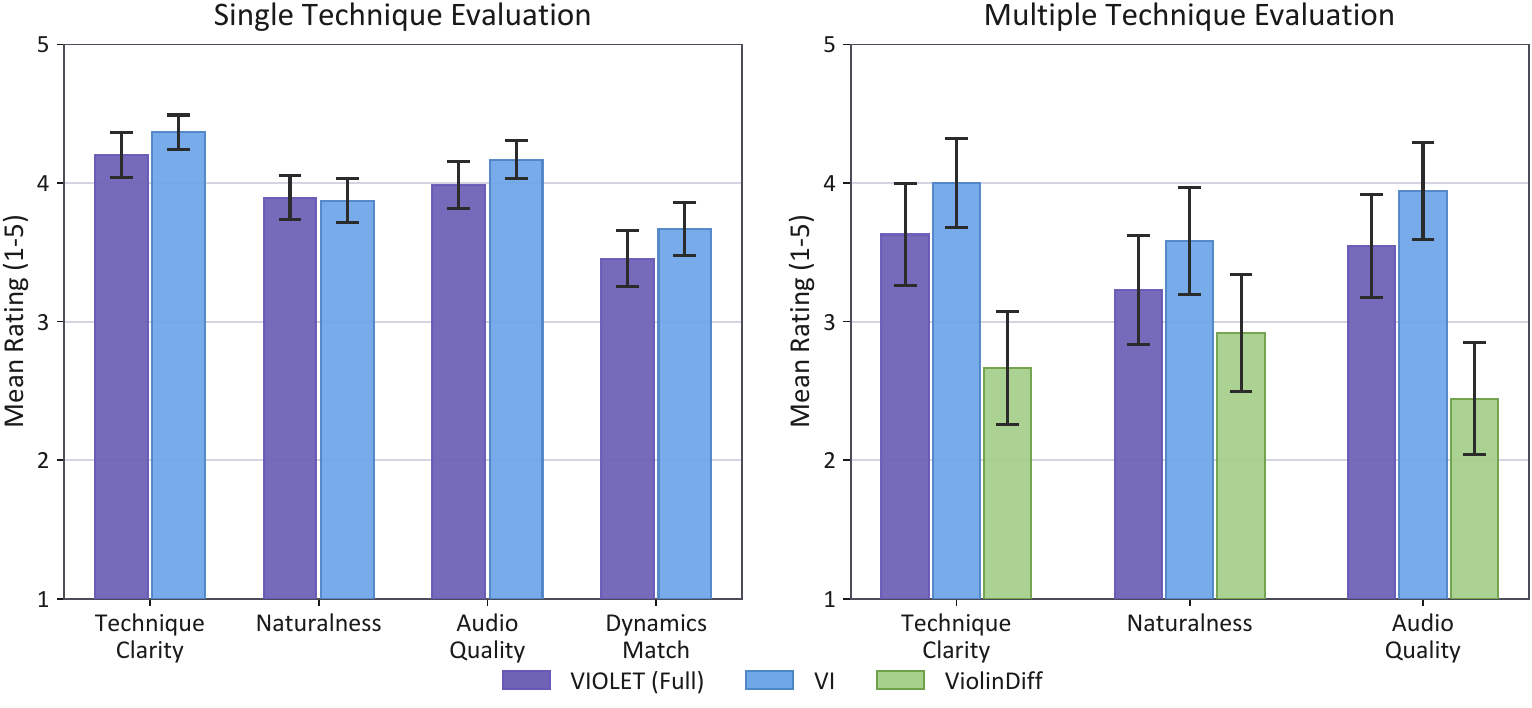}
  \caption{Mean single-technique (left) and multi-technique (right) ratings with 95\% confidence intervals. ViolinDiff appears only in the multi-technique setting.}

  \label{fig:subjective_rating}
  \vspace{-5mm}
\end{figure}

\Cref{fig:subjective_rating} reports the mean listener ratings. Statistical significance between systems is evaluated using a paired sign test under the null hypothesis that either system is equally likely to receive higher ratings. In the single-technique evaluation, VIOLET is comparable to the VI system in technique clarity ($p = 0.051$) and naturalness ($p = 1.000$), surpassing or reaching an average rating of 4. VIOLET slightly underperforms the VI system on audio quality ($p < 0.05$) and dynamics matching ($p < 0.01$), which aligns with our objective findings.

In the multi-technique evaluation, VIOLET receives significantly higher mean ratings than ViolinDiff in technique clarity and audio quality (both $p < 0.001$). The difference in naturalness is not statistically significant ($p = 0.122$). This is not surprising as ViolinDiff neither accepts technique or dynamics controls nor was retrained on our data, and it renders audio at 16 kHz. Nevertheless, this is still remarkable progress on neural audio synthesis for the violin. This result suggests that explicit technique conditioning improves perceptual quality in passages with multiple technique changes. Compared with the VI system, although VIOLET receives slightly lower ratings, the differences are not statistically significant in technique clarity ($p = 0.152$), naturalness ($p = 0.134$), and audio quality ($p = 0.053$). This demonstrates that VIOLET can successfully maintain performance stability and handle technique transitions on par with the VI baseline during long-form generation. 
Among the training data of VIOLET (Full), only CSV-TD training set contains both techniques and dynamics annotations, but the audio was rendered using the VI system and the technique assignments during rendering were not musically designed. These factors may explain the current limitations of VIOLET but also suggest promising directions in developing high-quality training data. 

\section{Conclusion}
In this paper, we presented \textbf{VIOLET}, a high-quality, controllable violin synthesis framework, together with \textbf{CSV-TD}, a new 48~kHz violin solo performance dataset with time-aligned MIDI notes, technique labels, and continuous dynamics curves. The proposed latent diffusion generation system renders violin audio with explicit control over both techniques and dynamics while preserving strong pitch and timing alignment. Objective and subjective evaluations show that it substantially improves over the state-of-the-art neural synthesis method and approaches a top commercial virtual instrument. To our best knowledge, this is the first music generative system that takes time-aligned conditioning of notes, playing techniques and continuous dynamics. 
For future work, we will scale the training data to online violin solo recordings to further enhance the expressiveness and naturalness of the synthesized audio. We will also move beyond explicit technique specifications toward automatic technique selection based on the musical context.

\section{AI Usage Statement}
During the preparation of this work, the authors utilized AI-assisted technologies to support both model development and paper preparation. For the coding and implementation phase, Cursor, OpenAI Codex, and Google Gemini were used to assist in writing, refactoring, and debugging code. For the preparation of the manuscript, OpenAI ChatGPT and Google Gemini were employed to polish the English language and improve overall readability. The authors have extensively reviewed and meticulously edited all AI-generated content, and take full responsibility for the final content, originality, and integrity of the published work.

\section{Acknowledgments}
This research was partially supported by National Science Foundation grant No. 2222129. We thank Yang Yi for his help with batch synthesis of violin audio using a virtual instrument in Kontakt. We sincerely thank the 15 musicians who voluntarily participated in the subjective evaluation. We also thank the reviewers and the meta-reviewer for their constructive comments that helped improve the paper.

\bibliography{ISMIRtemplate}

%
%
%
%

\end{document}